\documentclass[11pt]{article}

\usepackage{amsfonts}
\usepackage{amsmath}
\usepackage{float}
\usepackage{graphicx}
\usepackage{mathrsfs}

\usepackage{verbatim}
\usepackage{lineno}

\begin{document}

\begin{center}

\textbf{\large Multifractal analysis of three-dimensional grayscale images: \\
Characterization of natural porous structures}

\vspace{0.75cm}

\normalsize Lorenzo Milazzo$^1$ and Radoslaw Pajor$^2$

\end{center}

\vspace{0.25cm}

\noindent {\small $^1$\emph{Edinburgh, UK \\ [-0.2ex]
\hspace*{0.05cm} email: lorenzo.milazzo@physics.org} \\ [0.2ex]
$^2$\emph{Nottingham, UK \\ [-0.2ex]
\hspace*{0.05cm} email: radoslaw.pajor@nottingham.ac.uk}} \\ [0.2ex]

\vspace{1.5cm}

\begin{center}

\textbf{Abstract}

\end{center}

\noindent A multifractal analysis (MFA) is performed on three-dimensional grayscale images associated with natural porous structures (soil samples). First, computed tomography (CT) scans are carried out on the samples to generate 3D grayscale images. Then, a preliminary analysis is conducted to evaluate key quantities associated with the porosity, such as void fraction, pore volume, connectivity, and surface area. Finally, the samples are successfully identified and separated into two different structure families by using the MFA. A new software (\texttt{Munari}) to carry out the MFA of 3D grayscale images is also presented.

\vspace{2.0cm}

\noindent \textbf{1. Introduction}

\vspace{0.5cm}

\noindent Several techniques can be used to obtain representations of complex structures through \emph{3D images}. Examples of 3D images are the computed tomography (CT) images. Typically, a 3D image is constituted by a stack of 2D images.

\vspace{0.25cm}

\noindent Characterization of natural porous structures can contribute to better understand the physical processes in soil, in particular the transport processes. In general, the properties of these structures and the variables associated with them exhibit spatial irregularity. If the irregularity on the distribution of a given variable remains statistically similar at different scales, the variable is assumed to be self-similar. Thus, exploring self-similarity and scale invariance in natural porous structures can provide insights into the nature of the spatial variability of the soil properties. \\
The aim of the present paper is to present the multifractal analysis (MFA) of 3D grayscale images of soil samples and, more generally, to explore the possibility to use this method for the classification of 2D/3D complex structures.

\vspace{1.7cm}

\noindent \textbf{2. Multifractal Analysis (MFA): Theory and Methods}

\vspace{0.5cm}

\noindent Geometric (mono)fractals are self-similar sets of points. Multifractals (or multifractal measures) are self-similar measures defined on specific set of points [Baveye \emph{et al.} 2008; Evertsz \emph{et al.} 1992; Falconer 2003]. In general, the former can be generated by additive processes and the latter by multiplicative cascade of random processes. \\
The \emph{generalized fractal dimension}~$D_{q}$, which is closely related to the \emph{R\'{e}nyi entropy} [R\'{e}nyi 1961], provides a direct measurement of the fractal properties of an object -- several values of the \emph{momentum order} $q$ correspond to well-known generalized dimensions, such as the \emph{capacity dimension} (\emph{box-counting dimension})~$D_{0}$, the \emph{information dimension}~$D_{1}$, and the \emph{correlation dimension}~$D_{2}$. The \emph{singularity spectrum}~$f(\alpha)$ provides information about the scaling properties of the structure [Halsey \emph{et al.} 1986, Hentschel \emph{et al.} 1983; L\'{e}vy V\'{e}hel 1998; Lowen \emph{et al.} 2005; G. Paladin \emph{et al.} 1987; Theiler 1990].

\vspace{0.25cm}

\noindent The \emph{generalized fractal dimension} is defined by:

\begin{equation}
D_{q} = \lim_{l \to 0} \frac{1}{q-1} \frac{\ln{\sum_{i=1}^{N(l)} p_{i}^{q}(l)}}{\ln{l}} \label{DqEq}
\end{equation}

\vspace{0.25cm}

\noindent where $p_{i}(l)$ is the integrated measure associated with the \emph{i-th} box, $q$ is the momentum order, and $N(l)$ is the number of boxes of linear size $l$. The integrate measure is the concentration of the variable of interest in a given box relative to the whole system and it represents a probability -- \emph{e.g.} in the case of the mass, the integrate measure is the mass probability. \\
Once $D_{q}$ is known, the \emph{singularity spectrum} $f(\alpha)$ can be evaluated via a Legendre transformation:

\begin{equation}
f(\alpha(q)) = q \alpha(q) - \tau(q), \hspace*{1.0cm} \alpha(q) = \frac{d\tau(q)}{dq}
\end{equation}

\noindent where $\alpha$ is the \emph{singularity strength} and $\tau(q) = (q-1)D_{q}$ [Halsey \emph{et al.} 1986]. 

\vspace{0.25cm}

\noindent The singularity spectrum can also be directly (without knowing $D_{q}$) evaluated by using the method proposed by Chhabra \emph{et al.} [1989]. The first step of this approach consists of defining a family of normalized measures $\mu(q)$:

\begin{equation}
\mu_{i}(q,l) = \frac{[p_{i}(l)]^{q}}{\sum_{j=1}^{N(l)}[p_{j}(l)]^{q}} \label{def_mu}
\end{equation}

\vspace{0.25cm}

\noindent For each box~\emph{i}, the normalized measure $\mu_{i}(q,l)$ depends on the order of the statistical moment and on the box size and it takes values in the range [0,1] for any value of $q$. \\
Then, the two functions $f(q)$ and $\alpha(q)$ are evaluated:

\begin{eqnarray}
f(q)      & = & \lim_{l \to 0} \frac{\sum_{i=1}^{N(l)}\mu_{i}(q,l) \ln \mu_{i}(q,l)}{\ln l} \\ \label{def_sspectrum}
\alpha(q) & = & \lim_{l \to 0} \frac{\sum_{i=1}^{N(l)}\mu_{i}(q,l) \ln p_{i}(l)}{\ln l} \label{def_sstrength}
\end{eqnarray}

\vspace{0.25cm}

\noindent where $\alpha(q)$ is the \emph{average} value of the singularity strength
$\alpha_{i} = \ln p_{i}(l) / \ln l$. For each $q$, values of $f(q)$ and
$\alpha(q)$ are obtained from the slope of plots of $\sum_{i=1}^{N(l)}\mu_{i}(q,l) \ln \mu_{i}(q,l)$ versus
$(\ln l)$ and $\sum_{i=1}^{N(l)}\mu_{i}(q,l) \ln p_{i}(l)$ versus $(\ln l)$
over the entire range of box size values under consideration. Finally, the singularity spectrum $f(\alpha)$ is constructed from these two data sets.

\vspace{1.7cm}

\noindent \textbf{3. Materials and Methods}

\vspace{0.5cm}

\noindent The samples considered in this study are part of a larger set used by Harris \emph{et al.} [2003] and Pajor \emph{et al.} [2010] to analyse fungal colony spread through soil pore space. \\
The soil used to prepare the samples was a sandy loam soil from an experimental site at the Scottish Crop Research Institute (now James Hutton Institute), Invergowrie Dundee~UK. First, the material was air-dried and sieved to obtain aggregates with diameter of 1-2~mm; then, it was sterilized and packed into PVC rings with densities ranging from 1.2~g/cm\textsuperscript{3} to 1.6~g/cm\textsuperscript{3}.

\vspace{0.25cm}

\noindent The 3D volumes representing the soil samples were obtained by using the X-Tek (Metris) X-ray microtomography system. All samples were scanned with the same settings at 160~kV, 201~$\mu$A, with 0.1~mm Al filter in front of the X-ray gun with tungsten target. In total, 3003 angular projections (2D radiographs) were collected at 4 frames per second. The 2D radiographs were reconstructed into a 3D volume by using the \texttt{CT Pro} software; to perform this operation, a voxel size of 30~$\mu$m was chosen. The reconstructed volumes were then rendered and converted to a stack of grayscale TIFF images by using the \texttt{VG Studio Max} software. By using the \texttt{Fiji} software [Schindelin \emph{et al.} 2012], the 3D images (stacks of single-voxel thick, 8-bit images) were first cropped to $128\times128\times128$ voxels and then treated with median filter (radius = 2.0) and thresholded via a procedure based on the Ridler-Calvard method [Ridler \emph{et al.} 1978]. Finally, 3D grayscale images in ASCII format were generated from the stacks of TIFF images.

\vspace{0.25cm}

\noindent The measurements of void fraction, pore volume, connectivity, and surface area of the material were carried out by using the \texttt{Fiji/ImageJ} software and its plugin \texttt{BoneJ} [Doube \emph{et al.} 2010]. If an image is in \emph{grayscale} format, the pixel values lay in the range [0,255] and can be considered as measures of mass (0 = black = no mass; 255 = white = mass). Since the porous structures under analysis are represented by grayscale images, we also introduced the following metric $\phi^{gs}$ to characterize the porosity of the system:

\begin{equation}
\phi^{gs} = 1-((\text{sum of pixel values})/\text{sum of pixel values if no void in the sample})) \label{PorEq}
\end{equation}

\vspace{0.25cm}

\noindent It is important to highlight the limitations of this quantity: $\phi^{gs}$ is expected to be more effective in the case of grayscale images associated with high heterogeneous structures containing large pores.

\vspace{0.25cm}

\noindent A program -- \texttt{Munari} -- [Milazzo 2010] has been developed to perform MFA of 2D and 3D grayscale images and, in particular, to directly evaluate the singularity spectrum $f(\alpha)$ by using the Chhabra method. The application is written in~C++ and, at this stage, it processes grayscale images in ASCII format. The algorithms within \texttt{Munari} have been designed to be general and, as a result, they are independent from the resolution of the images and from the values of box sizes and moment orders used within the Chhabra method. Several validation tests have been implemented to ensure reliability and stability of the application; one of them is based on the analysis of 3D synthetic images representing multifractal lattices generated by using a random multiplicative process [Milazzo 2013]. \\
In this study, the \texttt{Munari} software was also used to evaluate the porosity-related metric~$\phi^{gs}$.

\vspace{1.7cm}

\noindent \textbf{4. Results and Discussion}

\vspace{0.5cm}

\noindent In this study, based on the analysis carried out by Pajor \emph{et al.} [2010], we focus on a set of eight samples. The samples S3DXX can be grouped into two families: four of them (`family 1': S3D01, S3D02, S3D03, S3D04) have a density equal to 1.3~g/cm\textsuperscript{3}, the other four (`family 2': S3D05, S3D06, S3D07, S3D08) a density equal to 1.6~g/cm\textsuperscript{3}. \\
There are significant differences in characteristics of pore geometry between the two families. Samples from `family 1' have a higher volume of pores that are thicker and better connected than those present in the structures grouped in `family 2', whereas samples from `family 2' have a higher surface area and higher number of micro and mesopores. \\
The measured values of porosity ($\phi$, [decimal fraction]), porosity-related metric ($\phi^{gs}$, [decimal fraction]), pore volume ($V_{V}$, [mm\textsuperscript{3}]), connectivity ($C$, [\%]), and surface area ($S$, [mm\textsuperscript{2}]) for the samples S3DXX are shown in Table~\ref{Table1} and ~\ref{Table2}.  \\
Note that, because of how it is defined, $\phi^{gs}$ is more effective in characterizing the porous structures of the `family 1' than those of the `family 2'.

\vspace{0.5cm}

\begin{table}[H]
\begin{center}
\begin{tabular}{@{\hspace{0.5cm}} l @{\hspace{0.5cm}} c
                @{\hspace{0.5cm}} c @{\hspace{0.5cm}} c
                @{\hspace{0.5cm}} c @{\hspace{1.0cm}} c
                @{\hspace{0.5cm}}}
\hline \hline
\\ [-1.0ex]
 \multicolumn{6}{c}{family 1} \\
\\ [-1.5ex]
\hline
 & S3D01 & S3D02 & S3D03 & S3D04 & mean($\sigma$) \\
\hline
\\ [-1ex]
 $\phi$     &  0.30 &  0.31 &  0.19 &  0.40 &  0.30(0.09) \\
 $\phi^{gs}$ &  0.75 &  0.77 &  0.73 &  0.80 &  0.76(0.03) \\
 $V_{V}$     & 13.43 & 13.67 &  8.45 & 17.57 & 13.28(3.74) \\
 $C$         & 97.42 & 98.42 & 95.53 & 96.30 & 96.92(1.27) \\
 $S$         & 35.25 & 27.73 & 25.92 & 31.53 & 30.11(4.15) \\
\\ [-1ex]
\hline \hline
\end{tabular}
\caption{\small{Key quantities associated with the \emph{porosity} for the samples S3D01-04.} \label{Table1}}
\end{center}
\end{table}

\vspace{0.5cm}

\begin{table}[H]
\begin{center}
\begin{tabular}{@{\hspace{0.5cm}} l @{\hspace{0.5cm}} c
                @{\hspace{0.5cm}} c @{\hspace{0.5cm}} c
                @{\hspace{0.5cm}} c @{\hspace{1.0cm}} c
                @{\hspace{0.5cm}}}
\hline \hline
\\ [-1.0ex]
 \multicolumn{6}{c}{family 2} \\
\\ [-1.5ex]
\hline
 & S3D05 & S3D06 & S3D07 & S3D08 & mean($\sigma$) \\
\hline
\\ [-1ex]
 $\phi$     &  0.15 &  0.08 &  0.14 &  0.10 &  0.12(0.03) \\
 $\phi^{gs}$ &  0.65 &  0.65 &  0.67 &  0.65 &  0.66(0.01) \\
 $V_{V}$     &  6.90 &  3.70 &  6.15 &  4.64 &  5.35(1.45) \\
 $C$         & 85.58 & 69.13 & 82.14 & 81.20 & 79.51(7.17) \\
 $S$         & 39.89 & 26.74 & 34.96 & 28.50 & 32.52(6.05) \\
\\ [-1ex]
\hline \hline
\end{tabular}
\caption{\small{Key quantities associated with the \emph{porosity} for the samples S3D05-08.} \label{Table2}}
\end{center}
\end{table}

\vspace{0.5cm}

\noindent MFA is one of the methods used to characterize and, in particular, to \emph{classify} natural porous structures. The characterization can be performed on either binary or grayscale images of the system [Dathe \emph{et al.} 2006; Lafond \emph{et al.} 2012; Posadas \emph{et al.} 2003; San Jos\'{e} Mart\'{i}nez \emph{et al.} 2010; Tarquis \emph{et al.} 2009; Zhou \emph{et al.} 2011]. The classification can be achieved by analysing the curves associated with either the singularity spectrum $f(\alpha)$ or the generalized fractal dimension $D_{q}$. In general, by using this approach, two or more groups of complex structures can be classified according the following main features: \emph{a)}~the width of the spectrum ($\alpha_{max} −- \alpha_{min}$); \emph{b)}~the position of the maximum of the spectrum; and \emph{c)}~the ratio between the information dimension and the capacity dimension, $D_{1}/D_{0}$ [Biswas \emph{et al.} 2012; Reljin \emph{et al.} 2002].

\vspace{0.25cm}

\begin{figure}[ht]
\rotatebox{-90}{
\centering
\includegraphics[scale=0.40]{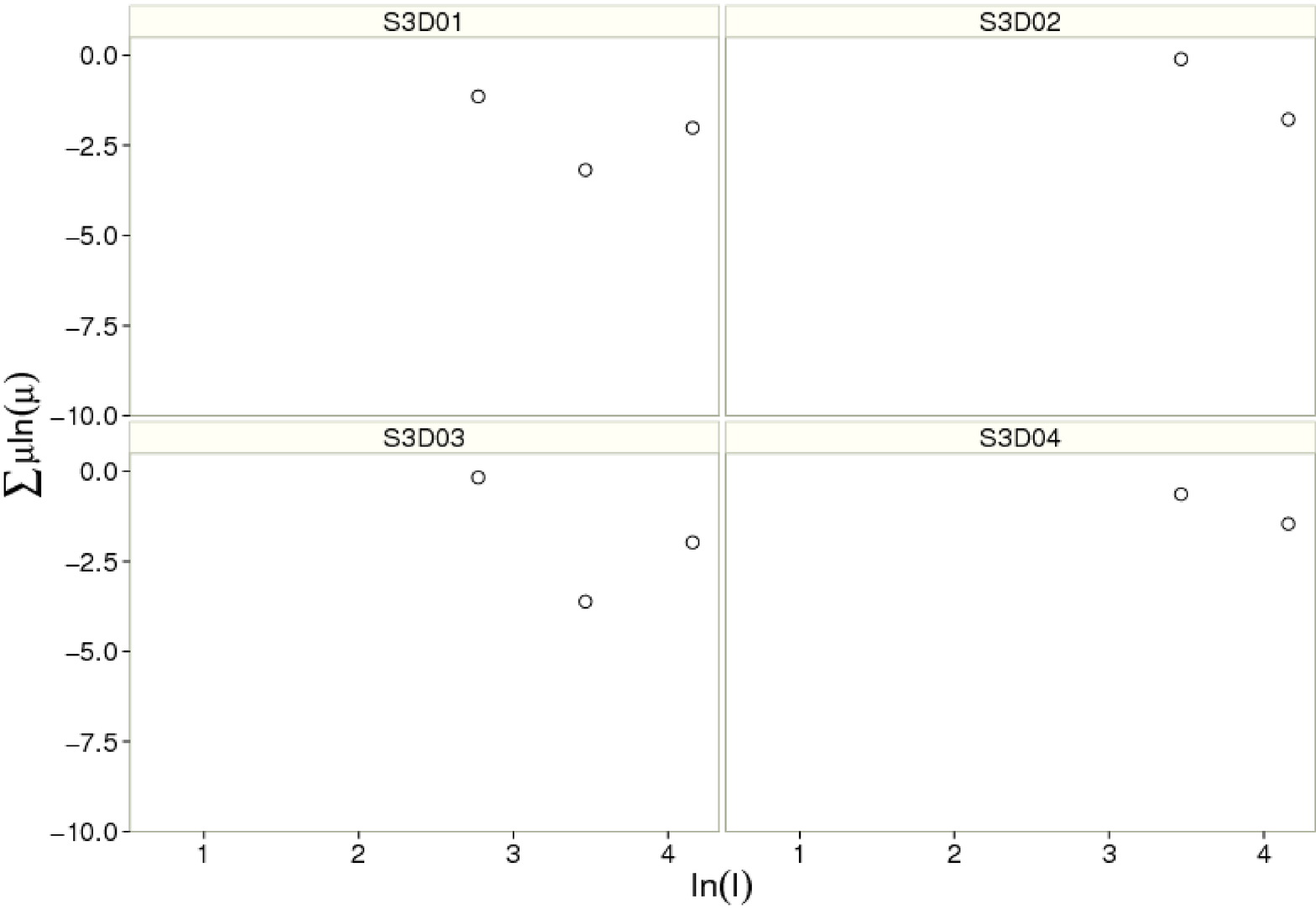}
}
\vspace*{-0.5cm}%
\caption{\small{Study of scaling behaviour: plots of $\sum_{i=1}^{N(l)}\mu_{i}(q,l) \ln \mu_{i}(q,l)$ vs $(\ln l)$ for $q=-3$. \newline Fractality is not detectable -- no linear fitting.} \label{f01}}

\vspace*{1.25cm}

\rotatebox{-90}{
\centering
\includegraphics[scale=0.40]{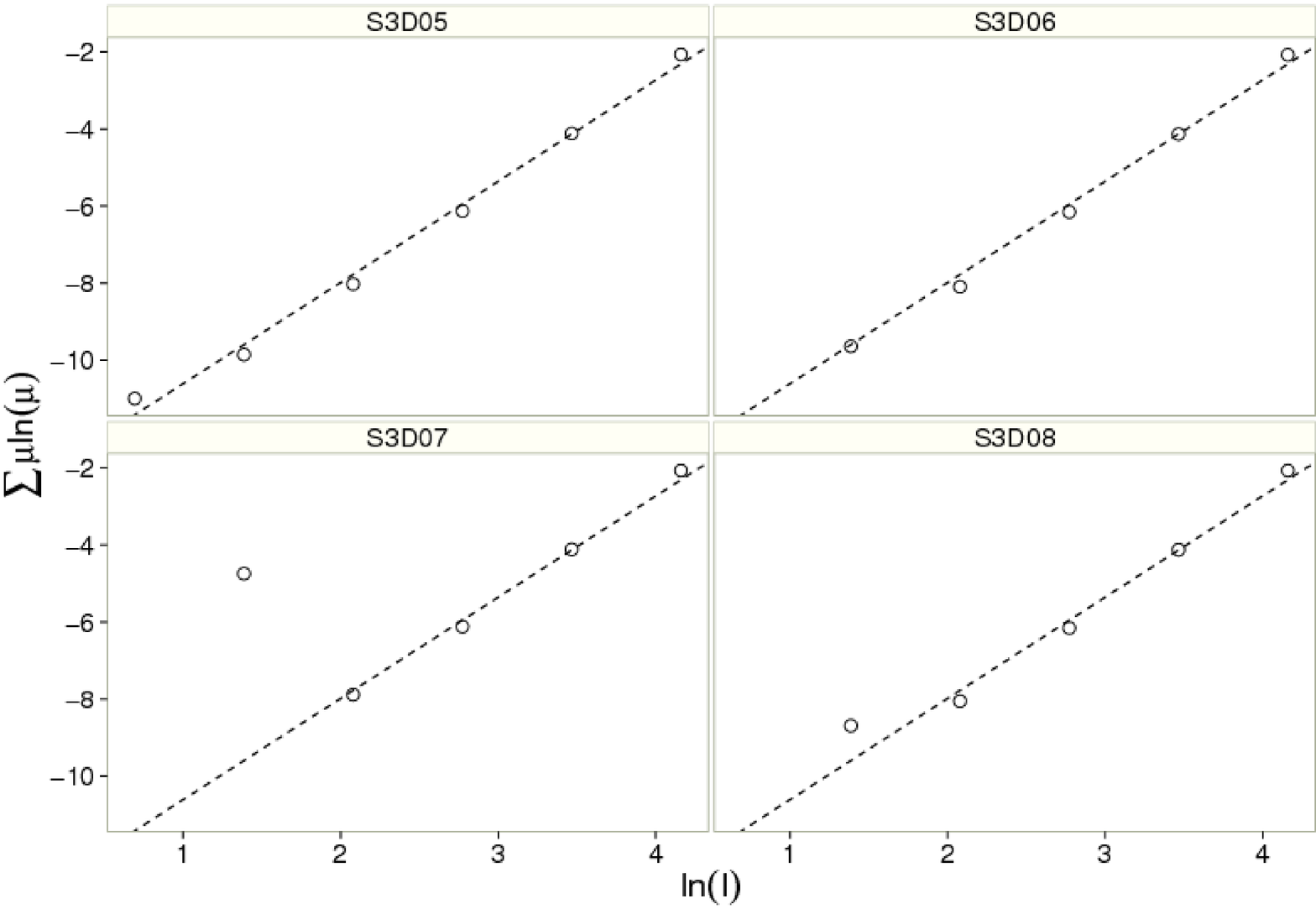}
}
\vspace*{-0.5cm}%
\caption{\small{Study of scaling behaviour: plots of $\sum_{i=1}^{N(l)}\mu_{i}(q,l) \ln \mu_{i}(q,l)$ vs $(\ln l)$ for $q=-3$. \newline Fractality is observed -- the slope of the fitting lines is ~$f(q=-3)$.} \label{f02}}
\end{figure}

\begin{figure}[t]
\rotatebox{-90}{
\centering
\includegraphics[scale=0.40]{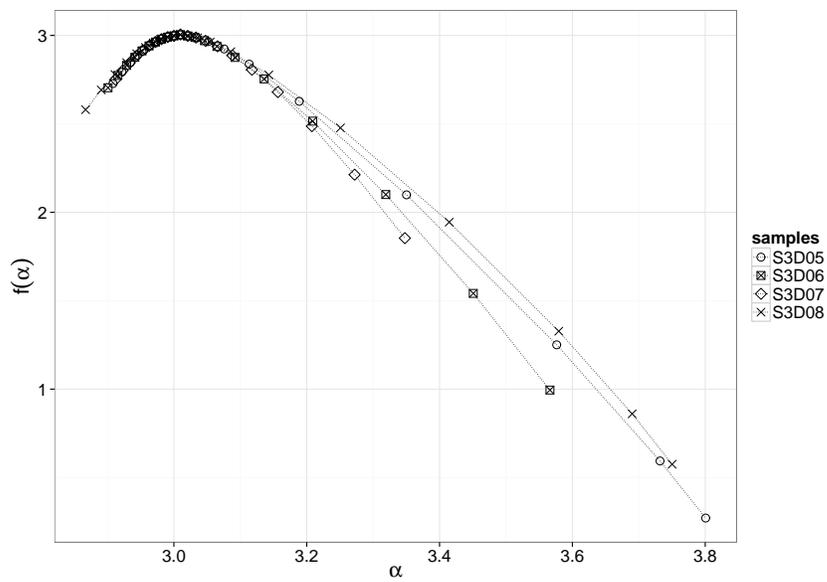}
}
\caption{\small{ -- Singularity spectra for the samples S3D05-08. \newline The maximum of $f(\alpha)$ is at $q=0$ and $\alpha=3.01$.} \label{f03}}
\end{figure}

\noindent In this study, we focus on the singularity spectrum and on its width. The \emph{singularity strength}~$\alpha$ is a local scaling index and the \emph{singularity spectrum}~$f(\alpha)$ represents the frequency of the occurrence of a certain value of the singularity strength -- in other words, $f(\alpha)$ captures how frequently a value of the local scaling index is found [Halsey \emph{et al.} 1986; G. Paladin \emph{et al.} 1987; Theiler 1990]. Moreover, $\alpha$ corresponds to the asymptotic behaviour of the \emph{coarse singular exponent} (\emph{coarse H\"{o}lder exponent}) $\alpha = \ln \mu / \ln l$ [Evertsz \emph{et al.} 1992]; this exponent represents the crowding or the degree of concentration of the measure: the greater this value is, the smaller is the concentration of the measure and vice versa [San Jos\'{e} Mart\'{i}nez \emph{et al.} 2010]. The \emph{width of the spectrum} ($\alpha_{max} −- \alpha_{min}$) is related to the heterogeneity of the local scaling index~$\alpha$; in the case of porous structures, it provides information on the scaling diversity associated with the distribution of the pores. \\
Each 3D image (image stack) represents a finite volume of linear size $L=128$~[pixels]. The image stacks were processed by using six values of box sizes: $l = \{2, 4, 8, 16, 32, 64\}$. For these complex structures, we evaluated the singularity spectrum $\alpha(q)$ by adopting the set of 21 values of momentum orders: $q~=~\{-5.0, -4.5, -4.0, \dots,$ $4.0, 4.5, 5.0\}$. Note that, for $q > 1$, the normalized measure $\mu(q)$ (Eq.~\ref{def_mu}) amplifies the more singular regions of the measure; while for $q < 1$, it amplifies the less singular regions [Chhabra \emph{et al.} 1989; San Jos\'{e} Mart\'{i}nez \emph{et al.} 2010]. We also analysed the scaling behaviour for various values of~$q$. Fig.~\ref{f01} and ~\ref{f02} show the quantity $\sum_{i=1}^{N(l)}\mu_{i}(q,l) \ln \mu_{i}(q,l)$ versus $(\ln l)$ for $q=-3$. \\
In the case of the samples S3D01-04 (Fig.~\ref{f01}), most of the normalized measures $\mu(q)$ are undefined for $q<0$; as a result, a fractal behaviour is not detectable by using this method and the singularity spectra were not evaluated. Conversely, fractality is observed in the case of the samples S3D05-08 (Fig.~\ref{f02}). \\
In Fig.~\ref{f02}, a degradation of the scaling can be noted for the sample S3D07; this effect, especially for negative values of $q$, has been already observed by other authors [Chhabra \emph{et al.} 1989; Dathe \emph{et al.} 2006]. Thus, for this sample, we reduced the scaling range under analysis in order to reduce the standard error of the regression (Chhabra method) and to improve the reliability of the singularity spectrum. \\ 
The singularity spectra for the samples S3D05-08 are shown in Fig.~\ref{f03}. As expected, the curves $f(\alpha)$ are convex with a single maximum at $q=0$ ($\alpha=3.01$) and with infinite slope at $q=\mp\infty$ [Halsey \emph{et al.} 1986]. The four samples have almost the same width of the spectrum ($\alpha_{max} −- \alpha_{min}$), suggesting that there are similarities in their distributions of the pores. \\
Thus, the initial classification (`family 1' and `2') is confirmed by the MFA: \emph{a)} the two groups of samples are different because fractality is observed only for the second group; \emph{b)} the samples from family 2 are characterized by similar singularity spectra.

\vspace{1.7cm}

\noindent \textbf{5. Conclusions}

\vspace{0.5cm}

\noindent The aim of this study has been to investigate the use of the MFA as a tool for characterization of complex structures. To our knowledge this is the first MFA performed to 3D porous structures, based on 3D grayscale images. In the case of grayscale images, all the studies that we are aware of have been conducted on 2D cross-sections of the samples (2D images).

\vspace{0.25cm}

\noindent Eight soil samples have been selected for CT scan imaging. Afterward, by using the MFA, they have been successfully identified and separated into two different structure families. \\
In order to carry out the MFA of 3D grayscale images, a new software for image processing (\texttt{Munari}) has been developed. \\
Several key implications have emerged from the analysis. In general, a degradation of the scaling is expected to be observed, especially for negative values of the momentum order. If this is the case, it is recommended to carefully choose the scaling range used for the calculations. A better study of the scaling could be performed and more reliable singularity spectra could be obtained if 3D images at higher resolution are processed and analysed -- however, note that this would introduce considerable challenges for the data management due to the number and size of the image files. Finally, the choice of a larger number of different types of structures to be compared would allow to further explore the potentialities of the MFA as a classification method. 

\vspace{1.7cm}

\noindent \textbf{Acknowledgements}

\vspace{0.5cm}

\noindent The soil samples were collected from an experimental site at the Scottish Crop Research Institute (now James Hutton Institute), Invergowrie Dundee UK. The 3D grayscale images were acquired by using the CT scanner at the SIMBIOS Centre, Abertay University, Dundee UK.

\vspace{1.7cm}

\noindent \textbf{Author Contributions}

\vspace{0.5cm}

\noindent Conceived and designed the study: \emph{LM} and \emph{RP}. Prepared the samples: \emph{RP}. Performed the CT scanning and image generation: \emph{RP}. Analyzed the data: \emph{LM} and \emph{RP}. Designed and developed the \texttt{Munari} software used for the MFA: \emph{LM}. Performed the MFA: \emph{LM}. Wrote the paper: \emph{LM} and \emph{RP}.

\newpage

\noindent \textbf{References}

\vspace{1.5 cm}

\noindent
P. Baveye \emph{et al.}, Introduction to fractal geometry, fragmentation processes and multifractal measures: Theory and operational aspects of their application to natural systems, in N. Senesi and K.J. Wilkinson~(Ed.), \emph{Biophysical Chemistry of Fractal Structures and Processes in Environmental Systems}, John Wiley \& Sons, Chichester, 11--67 (2008) \\ [1.0ex]
A. Biswas \emph{et al.}, Application of multifractal and joint multifractal analysis in soil science: A review, in S. Ouadfeul~(Ed.), \emph{Fractal analysis and chaos in Geosciences}, \\
InTech, Rijeka, 109--138 (2012) \\ [1.0ex]
\emph{BoneJ} (software), [http://bonej.org/] \\ [1.0ex] 
A.B. Chhabra \emph{et al.}, Direct determination of the $f(\alpha)$ singularity spectrum and its application to fully developed turbulence, \emph{Phys. Rev. A} 40, 5284--5294 (1989) \\ [1.0ex]
\emph{CT Pro} (software), Nikon Metrology NV, Leuven, Belgium \\ [1.0ex]
A. Dathe \emph{et al.}, Multifractal analysis of the pore- and solid-phases in binary two dimensional images of natural porous structures, \emph{Geoderma} 134, 318-326 (2006) \\ [1.0ex]
M. Doube \emph{et al.}, BoneJ: Free and extensible bone image analysis in ImageJ, \\
\emph{Bone} 47, 1076--1079 (2010) \\ [1.0ex]
C.J.G. Evertsz and B.B. Mandelbrot, Multifractal measures, in H.O. Peitgen \emph{et al.}~(Ed.), \emph{Chaos and Fractals: New Frontiers of Science}, \\
Springer-Verlag, New York, Appendix B, 921--953 (1992) \\ [1.0ex]
K.J. Falconer, \emph{Fractal Geometry: Mathematical Foundations and Applications}, \\
Wiley, Chichester (2003) \\ [1.0ex]
\emph{Fiji} (software), [http://fiji.sc/] \\ [1.0ex]
T.C. Halsey \emph{et al.}, Fractal measures and their singularities: The characterization of strange sets, \emph{Phys. Rev. A} 33, 1141--1151 (1986) \\ [1.0ex]
K. Harris \emph{et al.}, Effect of bulk density on the spatial organisation of the fungus Rhizoctonia solani in soil, \emph{FEMS Microbiol Ecol.} 44(1), 45--56 (2003) \\ [1.0ex]
H.G.E. Hentschel and I. Procaccia, The infinite number of generalized dimensions of fractals and strange attractors, \emph{Physica D}, 8, 435--444 (1983) \\ [1.0ex]
J.A. Lafond \emph{et al.}, Multifractal properties of porosity as calculated from computed tomography (CT) images of a sandy soil, in relation to soil gas diffusion and linked soil physical properties, \emph{Eur. J. Soil Sci.} 63, 861--873 (2012) \\ [1.0ex]
J. L\'{e}vy V\'{e}hel, Introduction to the multifractal analysis of images, in Y. Fisher~(Ed.), \emph{Fractal Image Encoding and Analysis}, Springer-Verlag, New York, 299-341 (1998) \\ [1.0ex]
S.B. Lowen and M.C. Teich, \emph{Fractal-Based Point Processes}, \\
Wiley-Interscience, New Jersey (2005) \\ [1.0ex]
L. Milazzo, \emph{Munari}: A program for multifractal characterization of complex systems~(2010) \\ [1.0ex]
L. Milazzo, Multifractal analysis of three-dimensional grayscale images: Estimation of generalized fractal dimension and singularity spectrum, \\
arXiv:1310.2719 [cond-mat.stat-mech] (2013) \\ [1.0ex] 
R. Pajor \emph{et al.}, Modelling and quantifying the effect of heterogeneity in soil physical conditions on fungal growth, \emph{Biogeosciences}, 7, 3731--3740 (2010) \\ [1.0ex]
G. Paladin and A. Vulpiani, Anomalous scaling laws in multifractal objects, \emph{Phys. Rep.}, 156(4), 147-225 (1987) \\ [1.0ex]
A.N.D. Posadas \emph{et al.}, Multifractal characterization of soil oore system, \emph{Soil Sci. Soc. Am. J.} 67, 1361--1369 (2003) \\ [1.0ex]
A. R\'{e}nyi, On measures of information and entropy, \emph{Proc. 4th Berkeley Sympos. Math. Statist. Probab.}, University of California Press, Berkeley, vol.1, 547--561 (1961) \\ [1.0ex]
I.S. Reljin and B.D. Reljin, Fractal geometry and multifractals in analyzing and processing medical data and images, \emph{Archive of Oncology} 10(4), 283--93 (2002) \\ [1.0ex] 
F. San Jos\'{e} Mart\'{i}nez \emph{et al.}, Multifractal analysis of discretized X-ray CT images for the characterization of soil macropore structures, \emph{Geoderma} 156, 32--42 (2010) \\ [1.0ex]
A.M. Tarquis \emph{et al.}, Pore network complexity and thresholding of 3D soil images, \emph{Ecological Complexity} 6, 230--239 (2009) \\ [1.0ex]
J. Schindelin \emph{et al.}, Fiji: An open-source platform for biological-image analysis, \emph{Nature methods} 9(7), 676--682 (2012) \\ [1.0ex]
T.W. Ridler and S. Calvard, Picture thresholding using an iterative selection method, \emph{IEEE Transactions on Systems, Man and Cybernetics} 8, 630--632 (1978) \\ [1.0ex]
J. Theiler, Estimating fractal dimension, \emph{J. Opt. Soc. Am. A} 7, 1055--1073 (1990) \\ [1.0ex]
H. Zhou \emph{et al.}, Multifractal analyses of grayscale and binary soil thin section images, \emph{Fractals} 19(3), 299--309 (2011) \\ [1.0ex]
\emph{VG Studio Max} (software), Volume Graphics GmbH, Heidelberg, Germany

\clearpage

%--
%\begin{comment}

\clearpage

%\end{comment}

\end{document}